\documentclass[sigconf]{aamas} 
\usepackage{balance} 

 \setcopyright{ifaamas}
 \acmConference[]{}{}{}
 \copyrightyear{}
 \acmYear{}
 \acmDOI{}
 \acmPrice{}
 \acmISBN{}

\acmSubmissionID{26}

\usepackage[boxed]{algorithm}
\usepackage[noend]{algorithmic}
\usepackage{booktabs}

\usepackage{flushend}
\usepackage{tikz}
\usetikzlibrary{decorations.pathreplacing,calligraphy}
\usetikzlibrary{arrows}

	\usepackage[normalem]{ulem}
\usepackage{nicefrac}
\usepackage{blkarray}

\usepackage{bm}
\usepackage{eucal}
\usepackage{verbatim}
\usepackage{tikz}
\usetikzlibrary{arrows}
\usepackage{todonotes}
\newcommand{\haris}[1]{\textcolor{blue}{\textbf{Haris says:} #1}}

\usepackage{enumerate}
			\usepackage{boxedminipage}
			\usepackage{xspace}

 \setcopyright{none}
 
 \usepackage{natbib}

\begin{document}

\setlength{\abovedisplayskip}{3pt}
\setlength{\belowdisplayskip}{3pt}

\title{Matching Algorithms under Diversity-Based Reservations}


%

\author{Haris Aziz}
\affiliation{%
  \institution{UNSW Sydney}
  \streetaddress{}
  \city{Sydney} 
  \state{} 
\country{Australia}
  \postcode{}
}
\email{haris.aziz@unsw.edu.au}

\author{Sean Morota Chu}
\affiliation{%
  \institution{UNSW Sydney}
  \streetaddress{}
  \city{Sydney} 
  \state{} 
\country{Australia}
  \postcode{}
}
\email{seanmorotachu@gmail.com}

\author{Zhaohong Sun}
\affiliation{%
  \institution{CyberAgent Inc.}
  \streetaddress{}
  \city{Tokyo} 
  \state{} 
\country{Japan}
  \postcode{}
}
\email{sunzhaohong1991@gmail.com}

%
%
%
%
%
%

  \begin{abstract}
	  Selection under category or diversity constraints is a ubiquitous and widely-applicable problem that is encountered in immigration, school choice, hiring, and healthcare rationing. 
These diversity constraints are typically represented by minimum and maximum quotas on various categories or types. We undertake a detailed comparative study of applicant selection algorithms with respect to the diversity goals.
  \end{abstract}


\maketitle

	
\section{Introduction}
How should we hire job applicants when we want to take both the overall merit as well as requirements of various departments into account? How should we decide on student intake while considering both entrance test scores and target numbers of scholarships for different categories? How should we ration healthcare resources when patients can avail {resources} under various categories?
Which applicants should be given an immigration slot when the government has targets for various categories? 
These fundamental and important questions constitute a recurring theme in allocation and selection decisions.
We consider a natural mathematical model for the problem that captures the main features of many of the problems discussed above. Although various choice rules and algorithms for selecting agents have been proposed, there has been little work carefully comparing the relative performance of these algorithms, especially from an experimental methodology. 
In this paper, we undertake one of the first detailed experimental studies to understand how well the algorithms perform with respect to capturing the intended diversity goals as well as selecting the highest priority applicants. We also try to understand the tradeoffs between merit and diversity.


We consider a very widely studied model of selection under diversity constraints.
Firstly, there is a baseline ordering over the applicants. The baseline ordering could be the merit ordering in the context of school admissions, or the need for treatment in the context of healthcare rationing. If no diversity constraints are present, the selection of agents is made only with respect to the baseline priority ordering. If the diversity constraints are additionally present, then both the priority ordering and the diversity constraints are used to make selection decisions. 

The diversity constraints or goals are represented by imposing minimum and maximum quotas on each of the types. In particular, {given one school $c$}, there is a lower quota of $q_{c,t}^1$ for the number of slots taken by agents for type $t$ and there is an upper quota of $q_{c,t}^2$ 
for the number of slots taken by agents for type $t$.
In the line of literature~(see, e.g.,  \citet{EHYY14a}) both lower and upper quotas are viewed as guidelines towards reaching diversity goals: firstly, fill up slots of those types whose minimum quotas have not been reached. As a secondary consideration, fill up slots of those types whose minimum quotas have been reached, but not their maximum quotas. 

Another feature of our setting is that applicants can satisfy multiple types such as being extra talented or being from a  {disadvantaged} group. Each applicant who is selected is assumed to count towards one of the types satisfied by them. Such a type could include a general public type. This way of accounting for representation has been referred to as the one-to-one convention, which is popular in Indian college admissions~\citep{SoYe19b}. Since we are not only interested in which agents are selected but also {in} how many target numbers of spots corresponding to relevant types are filled up, the output for our problems is not just a set of selected agents. Instead, it is a matching that matches each student to some type that the student satisfies. Such a matching not only gives information about the set of selected agents who are matched but also gives a count of how many seats of each type are used.

In this paper, we examine the following problem. 

\begin{quote}
\emph{In selection problems under minimum and maximum quota diversity goals, how do various algorithms perform with respect to satisfying diversity goals as well as merit?}	
\end{quote}

With respect to performance on merit, we will compare the outcomes of algorithms according to various objectives, including average rank, worst rank, and best rank. 
When considering diversity constraints captured by lower and upper diversity quotas, a natural question is how to {gauge} the level of diversity captured by a given set of applicants or a matching? A natural solution was provided by \citet{AzSu21a} who viewed each type $t$ as two ranks of slots corresponding {to} lower and upper quotas.
A set of agents provides \emph{maximal diversity} if there is a matching that matches the agents to the types in such a way that the number of rank 1 slots is maximized and given that the number of rank 2 slots is maximized.

One of the first algorithms for the problem was presented by \citet{EHYY14a} who assumed that each applicant can satisfy at most one type. The algorithm takes a natural greedy approach to first fill up slots corresponding to rank 1 and then to rank 2. It can suitably be extended to the case where agents may have multiple types. We will use the natural extension as one of the main algorithms whose performance we examine.  We will refer to the algorithm as EHYY.

%

Another algorithm that we consider is the \emph{horizontal choice rule} by \citet{SoYe20a} that was designed to optimally filling up seats when there is a single rank of slots. We consider two versions of the rule of \citet{SoYe20a}: SY1 optimizes the use of the first ranked slots and SY2 merges the first and second ranked slots and then optimizes the use of these slots.

\citet{AzSu21a} presented algorithms that achieve maximal diversity. We will refer to the algorithm as A-S. There are several other algorithms that have been proposed or are used in real-world systems. The goal of this paper is to undertake a {comparative} study of various {algorithms} for the problem and see how they fare in terms of maximal diversity. We check how various algorithms do in terms of filling up the first ranked slots. We also check how various algorithms do in filling up the first two ranks. 

From the specification, the A-S already maximizes the use of rank 1 slots and given that, it maximizes the use of rank 2 slots. One of the goals of the paper is to understand the extent to which it {performs} in relation to other existing approaches. We will also compare the algorithms with two baseline algorithms that {predominantly} care about the priority of the agents rather than diversity concerns. 




In this paper, {we} present several contributions. 
Firstly, we present {a} consistent specification of various algorithms for our setting with minimum and maximum quotas or equivalently rank 1 and rank 2 seats. Secondly, we perform one of the first experimental comparisons of prominent selection algorithms in achieving optimal diversity goals as well as average merit ranking of the agents. 
Next, we investigate the performance of prominent selection algorithms across a variety of different environments, thereby determining the environmental parameters affecting their performance. 

Some of the conclusions from the experiments include the following.
The total number of reserves and the selection capacity of a problem instance influence the performance of each algorithm.
As the number of reserves relative to selection capacity increases, the performance of diversity based algorithms is reduced with respect to satisfying merit compared to matching algorithms that ignore reserves.
When the total number of reserves is exceeded by the selection capacity, A-S and SY2 have equivalent performance, despite having different behaviour when total reserves exceed selection capacity.
Overall, A-S is the best algorithm at fulfilling reserves across two ranks but performs worse in selecting for merit compared to SY1 and SY2, which are optimal for filling the first and first two ranks of reserves respectively.
The performance of EHYY is close to optimality
on average when satisfying the first rank reserves, but its worst case performance is reduced when selection capacity and the number of reserves increase.

We find that, due to the various different characteristics of each algorithm, there is a necessary tradeoff between achieving merit and diversity goals, and the choice of algorithm can help negotiate between these two goals for any specific problem instance.

\section{Related Work}

The literature on matching under diversity and other distributional constraints is vast. We discuss work that is closely related to our problem. 

Affirmative action in two-sided matching has been considered in early work on school choice \citet{AbSo03b,Abdu05a}. In many of the diversity models, each school puts a minimum quota on each type \citep{HYY13a,Koji12a,KoSo13a,EHYY14a}.
\citet{EHYY14a}  treated the quotas in a soft manner since hard constraints can lead to infeasibility. We pursue the same approach as well. In {contrast} to \citet{EHYY14a}, we allow agents to have multiple types.

The issue of agents having multiple `overlapping types' has been considered in recent papers and deployed applications in the past few years, including those in Brazil, Chile, Israel, and India (see, e.g.,~\citep{AyTu16a,BCC+19a,CEE+19a,KHIY17a,GNKR19a}). There are two ways to perform accounting when agents have multiple types~\citep{SoYe20a}. In the \textit{one-for-all} convention, an agent is viewed as taking slots for all the types that they satisfy~\citep{GNKR19a,AGS20a}. In the \textit{one-for-one} convention, they take the slot of one of the types they satisfy. In this paper, we pursue the one-for-one convention. This convention has the `more widespread interpretation'~\citep{SoYe20a}. The one-for-one {convention} has been explicitly or {implicitly} considered in several recent papers~~\citep{AyTu16a,KHIY17a,BCC+19a,CEE+19a,EHYY14a}. Most of these approaches do not achieve diversity optimally. In {contrast}, \citet{AzSu21a} presented a rule that achieves diversity optimally. When there is only one rank of reserves or when there are no maximum quotas, \citet{SoYe20a} presented a rule that also satisfies diversity optimally. We will consider two extensions of the algorithm for our model.

%
%

\section{Preliminaries}

An instance $I$ of the problem consists of a tuple $(S,c,q_c,T,\succ_c, \eta_c)$
 where $S$ denotes the set of agents. {There is one school $c$ with capacity $q_c$.}
We denote by $T$ the set of types.
We overload the term to also capture the information about the types of each agent. For each agent $s$, let $T(s) \subseteq T$ denote the subset of types to which agent $s$ belongs. If $T(s) = \emptyset$, then agent $s$ does not have any privileged type. 
{We use the term $\eta_c$ to specify the diversity goals of school $c$. In this work, we consider two ranks of slots.} 
The term $\eta_{c,t}^1$ {denotes the} number of slots of rank 1 of type $t$ {(minimum quotas)}
and $\eta_{c,t}^2$ denotes slots of rank 2 of type $t$ {(maximum quotas)}.\

The school $c$ has a strict priority ordering $\succ_c$ over $S \cup \{\emptyset\}$ where $\emptyset$ represents the option of leaving seats vacant for school $c$. An agent $s$ is \emph{acceptable} to school $c$ if $s \succ_c \emptyset$ holds. 
The priority ordering of the school could be based on the entrance exam scores, or in the case of automated hiring, on some objective measure that captures the suitability of the applicants.

\begin{example}
	\label{ex:1}
	Consider the setting in which there are six students $S = \{s_1,s_2,s_3,s_4,s_5,s_6\}$, applying for seats at one school $c$.
	The type profile of the students is $T(s_1) = \{\}$, $T(s_2) = \{t_{4}\}$ , $T(s_3) = \{t_{3}\}$, $T(s_4) = \{t_{1},t_{2},t_{3}\}$, $T(s_5) = \{t_{1}\}$ , $T(s_6) = \{t_{2},t_{3}\}$.
	The capacity of the school is $q_c = 3$ and the school has diversity goals specified as follows: $\eta^1_{c,t_{1}} = 1, \eta^1_{c,t_{2}} = 1, \eta^2_{c,t_{3}} = 1, \eta^2_{c,t_{4}} = 1$.
	The priority ordering of students is $s_1 \succ_c s_2 \succ_c s_3 \succ_c s_4 \succ_c s_5 \succ_c s_6$.

The interpretation of the diversity goals outlined for school $c$ {is} as follows: school $c$ wishes to admit 3 students while matching as many students to slots of rank 1 as possible.
	In the event that no further rank 1 slots can be matched, school $c$ would like to match as many rank 2 seats as possible.
	School $c$ has one rank 1 slot each for types $t_{1},t_{2}$ and one rank 2 slot each for types $t_{3},t_{4}$.
\end{example}



\section{A Tool Box of Algorithms}
\label{sec:toolbox}

{In this section, we describe several algorithms that we considered in the experiments.}
Diversity goals have been defined over two ranks, in the sense that first rank diversity goals are to be satisfied before second rank diversity goals whenever possible.
This is analogous to diversity settings in which minimum quotas are to be satisfied as many as possible before targeting maximum quotas.
Within this setting, we also allow for overlapping types, such that any agent may be prescribed multiple undersubscribed types.

\subsection*{A-S algorithm of \citeauthor{AzSu21a}}

The A-S algorithm creates a ranked reservation graph and then computes a rank-maximal matching within this graph to find a matching which optimizes first rank seat usage before second rank usage.

	Given a set of students $S'$ and a school $c$ with reserved quotas $\eta_c$, 
	a corresponding \emph{ranked reservation graph} $G=(S' \cup V, E, \eta_c)$ 
	is a bipartite graph whose vertices consist of a set of students $S'$ and a set of reserved seats $V$. Each reserved seat $v_{t,i}^j \in V$ has a rank $j$, a type $t$ and an index $i$. For each rank $j$ and each type $t$, we create $\eta_{c,t}^j$ reserved seats in $G$. The edge set $E$ is specified as follows. There is an edge between a student $s$ and a reserved seat 
	$v_{t,i}^j$ if {student $s$ belongs to type $t$},
	i.e., $t \in T(s)$. 
	Each edge $(s, v_{t,i}^j)$ has a \emph{rank} $j$ corresponding to the rank $j$ of the reserved seat $v_{t,i}^j$. We refer to all edges with rank $j$ as $j$-ranked edges. Since we are focusing on problems arising from lower and upper quotas, we assume that there are two main ranks $1$ and $2$. We also create an artificial universal type $t_0$ that each agent is eligible for and that has zero seats of rank $1$ and $2$ but $q_c$ seats of rank $3$.  Only those agents are matched to this type, if they are unable to be matched to seats of the real types. This type is only present to match those agents who are unable to be matched to real types. 
	To keep our figures simple, we will not depict vertices corresponding to $t_0$.
		
		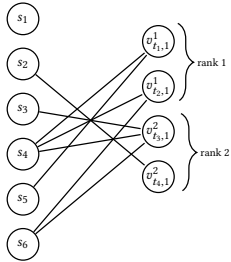
\begin{figure}[h!]
		
			\begin{center}
				  \scalebox{0.6}{
		\begin{tikzpicture}[-,>=stealth',shorten >=1pt,auto,node distance=3cm,
				thick,main node/.style={circle,fill=white!20,draw,minimum size=0.7cm,inner sep=0pt]}, scale=0.5]
	
			\node[main node] at (0, 3) (1)  {$s_1$};
			\node[main node] at (0, 1) (2)  {$s_2$};
			 \node[main node] at (0, -1) (3)  {$s_3$};
			   \node[main node] at (0, -3) (4)  {$s_4$};
			   \node[main node] at (0,-5) (5)  {$s_5$};
			   \node[main node] at (0,-7) (6)  {$s_6$};

		 \node[main node] at (6, 2) (d1)  {$v_{t_1,1}^1$};
		 \node[main node] at (6, 0) (d2)  {$v_{t_2,1}^1$};
	     \node[main node] at (6, -2) (d3)  {$v_{t_3,1}^2$};
	    \node[main node] at (6,-4)  (d4)  {$v_{t_4,1}^2$};
		 \draw [decorate,decoration={brace,amplitude=10pt},xshift=-4pt,yshift=0pt] (7,2.8) -- (7,-0.65) node [black,midway,xshift=9pt] {\footnotesize rank $1$};
	
		 \draw [decorate,decoration={brace,amplitude=10pt},xshift=0pt,yshift=0pt] (7,-1.2) -- (7,-4.65) node [black,midway,xshift=9pt] {\footnotesize rank $2$};

		\draw[] (2) -- node[above] {} (d4);
		\draw[] (3) -- node [above] {} (d3);
		\draw[] (4) -- node [above] {} (d1);
		\draw[] (4) -- node [above] {} (d2);
		\draw[] (4) -- node [above] {} (d3);
		\draw[] (5) -- node [above] {} (d1);
		\draw[] (6) -- node [above] {} (d2);
		\draw[] (6) -- node [above] {} (d3);
		%
		\end{tikzpicture}
		}
		\end{center}
		\caption{The ranked reservation graph for the problem instance in Example~\ref{ex:1}.}
		\label{fig:A-S-example1}
		\end{figure}

	\begin{algorithm}[h!]
	        \begin{algorithmic}[scale=1]
	        \REQUIRE $S'\subseteq S$, $q_c$, $\eta_c$, $\succ_c$.
	        \ENSURE A matching $M\subseteq S'\times V$ and 
			a set of matched agents $S^*\subseteq S'$
	        \end{algorithmic}
	      \begin{algorithmic}[1]
	      \caption{A-S algorithm from \citet{AzSu21a}}
	      \label{alg:choice_school1-for-1}
	       \STATE Selected agents $S^* \leftarrow \emptyset$ 
		  \STATE {Matching  $M \leftarrow \emptyset$}
		 \STATE Construct the corresponding ranked reservation graph $G=(S'\cup V, E, \eta_c)$. 
	\FOR{agent $s\notin S^*$ down the list in $\succ_c$}
	\IF{there exists a matching in $G$ of size at most $q_c$ that satisfies the following two conditions
	\begin{enumerate}
		\item it is rank maximal among all matchings in $G$ of size at most $q_c$
		\item it matches all agents in $S^*\cup \{s\}$
	\end{enumerate}}
	   \STATE Add $s$ to $S^*$
	   \ENDIF 
	   \ENDFOR
	   \STATE Compute a rank maximal matching $M$ of $G$ that matches all the students in $S^*$
	           \RETURN $M$ and $S^*$.
	       \end{algorithmic}
	       \end{algorithm}
	
\begin{figure}[h!]
		\begin{center}
			  \scalebox{0.6}{
	\begin{tikzpicture}[-,>=stealth',shorten >=1pt,auto,node distance=3cm,
			thick,main node/.style={circle,fill=white!20,draw,minimum size=0.7cm,inner sep=0pt]}, scale=0.5]
	
		\node[main node] at (0, 3) (1)  {$s_1$};
		\node[main node] at (0, 1) (2)  {$s_2$};
		 \node[main node] at (0, -1) (3)  {$s_3$};
		   \node[main node] at (0, -3) (4)  {$s_4$};
		   \node[main node] at (0,-5) (5)  {$s_5$};
		   \node[main node] at (0,-7) (6)  {$s_6$};
	
		\node[main node] at (6, 2) (d1)  {$v_{t_1,1}^1$};
		\node[main node] at (6, 0) (d2)  {$v_{t_2,1}^1$};
		\node[main node] at (6, -2) (d3)  {$v_{t_3,1}^2$};
	   \node[main node] at (6,-4)  (d4)  {$v_{t_4,1}^2$};
		\draw [decorate,decoration={brace,amplitude=10pt},xshift=-4pt,yshift=0pt] (7,2.8) -- (7,-0.65) node [black,midway,xshift=9pt] {\footnotesize rank $1$};
	   
		\draw [decorate,decoration={brace,amplitude=10pt},xshift=0pt,yshift=0pt] (7,-1.2) -- (7,-4.65) node [black,midway,xshift=9pt] {\footnotesize rank $2$};

	\draw[] (2) -- node[above] {} (d4);
	\draw[] (4) -- node [above] {} (d2);
	\draw[] (5) -- node [above] {} (d1);

	%
	\end{tikzpicture}
	}
	\end{center}
	\caption{The matching returned by the A-S Algorithm for the problem instance in Example~\ref{ex:1}.}
	%
	\label{fig:A-Sexample2}
	\end{figure}
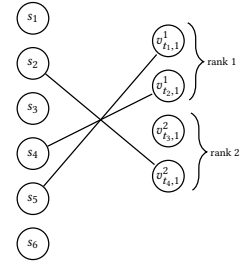

	\begin{example}[A-S Algorithm]
		\label{ex:A-S}
		For Example~\ref{ex:1}, 
		the A-S algorithm will select the students $S^{*} = \{s_2,s_4,s_5\}$, which fills both rank 1 slots and a rank 2 slot for type $t_4$.
		This arises from the ranked reservation graph $G$ pictured in Figure~\ref{fig:A-S-example1}.
		Evidently, the 5 possible rank maximal matchings on $G$ of size 3 are 
		$\{\{(s_2,v^{2}_{t_4,1}),(s_4,v^{1}_{t_2,1}),(s_5,v^{1}_{t_1,1})\}$,$\{(s_2,v^{2}_{t_4,1}),(s_4,v^{1}_{t_1,1}),(s_6,v^{1}_{t_2,1})\}$,
		$\{(s_3,v^{2}_{t_3,1}),(s_4,v{1}_{t_2,1}),(s_5,v^{1}_{t_1,1})\}$,$\{(s_3,v^{2}_{t_3,1}),(s_4,v^{1}_{t_1,1}),(s_6,v^{1}_{t_2,1})\}$,
		$\{(s_4,v^{2}_{t_3,1}),(s_5,v^{1}_{t_1,1}),(s_6,v^{1}_{t_2,1})\}\}$. 
		The A-S algorithm will first select $s_2$ when scanning the students by priority ordering, as there exists a rank maximal matching in $s_2$ is matched.
		It will then select $s_4$, as $s_3$ cannot be in the same rank-maximal matching as $s_2$, before finally selecting $s_5$.
		
		As the A-S algorithm selects the highest priority students possible while maintaining a rank-maximal matching, the final matching in this instance will be $\{(s_2,v^{2}_{t_4,1}),(s_4,v^{1}_{t_2,1}),(s_5,v^{1}_{t_1,1})\}$, as pictured in Figure~\ref{fig:A-Sexample2}. To keep our figure simple, we have not depicted vertices corresponding to $t_0$.

	\end{example}

	   \subsection*{EHYY Algorithm of \citeauthor{EHYY14a}}

	   One of the first algorithms for the problem was presented by \citet{EHYY14a} who assumed that each applicant can satisfy at most one type. When each agent has at most one type, the choice of which type's slot an agent should take is straightforward. The algorithm proposed by \citet{EHYY14a} follows a natural idea that provides the blueprint for many of the other algorithms in the literature. The algorithm works as follows. The algorithm goes down the priority list and selects the highest priority agent who has a type that is undersubscribed (whose count has not reached the lower quota). If there is no such agent, the algorithm selects agents with the highest priority who has some type that is not oversubscribed (whose count has not reached the upper quota). If there are no such agents, then the highest priority agents are selected until the total capacity is reached. When agents may have multiple types, the algorithm of \citet{EHYY14a} can be suitably generalized to handle `overlapping types'. When choosing an agent we select the highest priority agent who has some type that is undersubscribed; and otherwise we select the highest priority agents who have some type that is not oversubscribed. We will refer to the algorithm as EHYY.

The EHYY algorithm below approaches the two-ranked school choice problem by greedily selecting agents with first rank types before selecting agents with second rank types.
This approach is not optimal when there are multiple overlapping types for agents.
\begin{algorithm}[h!]
\begin{algorithmic}[scale=1]
	\REQUIRE $S'\subseteq S$, $q_c$, $\eta_c$, $\succ_c$.
    \ENSURE A matching $M\subseteq S'\times V$ and 
	a set of matched agents $S^*\subseteq S'$
	\end{algorithmic}
	\begin{algorithmic}[1]
	\caption{EHYY Algorithm of \citet{EHYY14a}}
	\label{alg:choice_school1-for-2}
	\STATE Selected agents $S^* \leftarrow \emptyset$ 
		  \STATE {Matching  $M \leftarrow \emptyset$}
	\FOR{agent $s\notin S^*$ down the list in $\succ_c$}
	\IF{ $|S^*| < q_c$ and if there exists an unmatched first rank seat $v_{t,i}^1$ for some type $t$ satisfied by $s$}
	\STATE Add $s$ to $S^*$ and $(s,v_{t,i}^1)$ to $M$
	\ENDIF 
	\ENDFOR
	\FOR{agent $s\notin S^*$ down the list in $\succ_c$}
	\IF{ $|S^*| < q_c$ and if there exists a second rank seat $(s,v_{t,i}^2)$ for some type $t$ satisfied by $s$}
	\STATE Add $s$ to $S^*$ and $(s,v_{t,i}^2)$ to $M$
	\ENDIF 
	\ENDFOR
	\FOR{agent $s$ down the list in $\succ_c$}
	\IF{$|S^*|<q_c$ and $s\notin S^*$}
	 \STATE Add $s$ to $S^*$ and add some new edge $(s,v_{t,i}^3)$ to $M$ where $t\in T(s)$.
	\ENDIF 
	\ENDFOR
	       \RETURN $M$ and $S^*$.
\end{algorithmic}
\end{algorithm}

\begin{example}[EHYY Algorithm]
	\label{ex:EHYY}
	Consider the problem instance described in Example~\ref{ex:1}.
	In this problem instance, EHYY may select $S^{*} = \{s_2,s_4,s_6\}$, filling two rank 1 slots and one rank 2 slot.

	In the first traversal of the priority list of students, we match students $s_4$ and $s_6$ to the unmatched slots $v^{1}_{c,t_1}$ and $v^{1}_{c,t_2}$ respectively.
	When matching $s_4$, we have two options: $v^{1}_{c,t_1}$ and $v^{1}_{c,t_2}$ - by random tiebreaking, we choose $v^{1}_{c,t_1}$. 
	
	In the second traversal of the priority list of students, we select student $s_2$ by matching her to the unmatched slot $v^{2}_{c,t_4}$.
	Our final matching is $\{(s_2,v^{2}_{t_4,1}),(s_4,v^{1}_{t_1,1}),(s_6,v^{1}_{t_2,1})\}$.
\end{example}

\subsection*{Horizontal Choice Algorithms of \citeauthor{SoYe20a}}

The horizontal choice algorithm was proposed by \citet{SoYe20a} for the case where a school has only one rank of reserves and once the reserves are filled up, the remaining seats are filled according to the priority ranking. 
We will refer to the algorithm as the SY algorithm. The algorithm gives the same outcome as the A-S algorithm for the case of one rank of reserves (along with the generic type $t_0$ that takes up any remaining agents who are not matched to reserves types) so we do not define it formally in the way  \citet{SoYe20a} did.
We adapt the SY algorithm from \citet{SoYe20a} by either focusing on one rank only, or by merging the two ranks of quotas into one.
The original algorithm SY does not allow for any type that has a rank $3$. However, we can view SY as having one type of rank 3 that matches all agents who were unable to be matched to an actual reserved seat. 
%

As a result of this preprocessing, we are testing two different algorithms: which we will call SY1 and SY2. 
SY1 below eliminates second rank seats from consideration when running the selection algorithm.
\begin{algorithm}[h!]
\begin{algorithmic}[scale=1]
        \REQUIRE $S'\subseteq S$, $q_c$, $\eta_c$, $\succ_c$.
        \ENSURE A matching $M\subseteq S'\times V$ and 
		a set of matched agents $S^*\subseteq S'$
        \end{algorithmic}
      \begin{algorithmic}[1]
      \caption{SY1 per \citet{SoYe20a}}
      \label{alg:choice_school1-for-3}
		\STATE Eliminate second rank seats from $\eta_c$ to get $\eta_c'$
		\STATE Run the horizontal choice rule of \citet{SoYe20a} with respect to $\eta_c'$; 
{let $M$ denote the}
matching returned by the horizontal choice rule 
	       \RETURN $M$ and $S^*$ (the set of agents matched by $M$).
       \end{algorithmic}
       \end{algorithm}

	   \begin{example}
		\label{ex:SY1}
		Consider the problem instance described in Example \ref{ex:1}. 
		In this instance, SY1 will select students $S^{*} = \{s_1,s_4,s_5\}$, 
		filling two rank 1 slots and no rank 2 slots. 
		This is as we remove all rank 2 slots from our instance, so the remaining slots are $v^{1}_{t_1,1}$ and $v^{1}_{t_2,1}$.
		Any matching that includes these two slots will be rank maximal.
		
		In order to fill the two rank 1 slots, SY1 will pair $s_4$ with $v^1_{t_2,1}$, and $s_5$ with $v^{1}_{t_1,1}$. 
		Due to the lack of second rank seats, SY1 will then select the highest priority student $s_1$, pairing $s_1$ with $v^{3}_{t_0,1}$, where $t_0$ is a generic type shared by all students.
		Hence the final matching for SY1 will be $\{(s_4,v^1_{t_2,1}),(s_5,v^{1}_{t_1,1}),(s_1,v^{3}_{t_0,1})\}$.

	   \end{example}

SY2 below merges first and second rank seats into seats for a single rank, such that both ranks are considered at the same priority.
\begin{algorithm}[h!]
\begin{algorithmic}[scale=1]
		\REQUIRE $S'\subseteq S$, $q_c$, $\eta_c$, $\succ_c$.
        \ENSURE A matching $M\subseteq S'\times V$ and 
		a set of matched agents $S^*\subseteq S'$
		\end{algorithmic}
		\begin{algorithmic}[1]
		\caption{SY2 per \citet{SoYe20a}}
		\label{alg:choice_school1-for-4}
		\STATE Merge first and second rank seats from $\eta_c$ into one rank to get $\eta_c'$
		\STATE Run the horizontal choice rule as per \citet{SoYe20a} with respect to $\eta_c'$
    \STATE $M \leftarrow$ matching returned by the horizontal choice rule 
       \RETURN $M$ and $S^*$ (the set of agents matched by $M$).
		\end{algorithmic}
		\end{algorithm}

		\begin{example}[SY2 Algorithm]
			\label{ex:SY2}
			Consider the problem instance described in Example \ref{ex:1}.
			Since SY2 considers all seats as equal rank, for this instance any matching where all students are matched to a ranked seat is considered rank maximal.
			Hence, SY2 will select students $S^{*} = \{s_2,s_3,s_4\}$, filling one rank 1 slot and two rank 2 slots.
			This is as, we ignore $s_1$ due to a lack of types, then match $s_2,s_3,s_4$ sequentially as they can be matched as $\{(s_2,v^2_{t_4,1}),(s_3,v^{2}_{t_3,1}),(s_4,v^{1}_{t_1,1})\}$, which is our final matching.
			Other rank-maximal matchings, in this case, are ignored as they require selecting students of lower rank in the priority list.
		\end{example}



\subsection*{Priority Only Algorithms}

Next, we discuss two algorithms that select agents only on the basis of their priority. In other words, they select the top $q_c$ agents. Since we are not only interested in the selection of agents but also want to check the type used by each selected student, the algorithms return matchings {by specifying which student is matched with which slot}. 

The first priority only algorithm Priority Only Greedy (POG) goes down the priority list and for a current agent, gives them a rank 1 seat from an eligible type, and if such seat is not available, then a rank 2 seat from an eligible type. If neither of the two ranks are available, then a rank 3 seat from an eligible type is matched to the agent. 

The second priority only algorithm selects the same set of agents but matches them to ranked seats in a smart way. For this reason, we refer to it as Priority Only Smart (POS). The algorithm uses A-S to match the set of selected students in an optimal way to the ranked seats.

\begin{algorithm}[h!]
\begin{algorithmic}[scale=1]
	\REQUIRE $S'\subseteq S$, $q_c$, $\eta_c$, $\succ_c$.
    \ENSURE A matching $M\subseteq S'\times V$ and 
	a set of matched agents $S^*\subseteq S'$
	\end{algorithmic}
	\begin{algorithmic}[1]
	\caption{Priority Only Greedy (POG)}
	\label{alg:pog}
	\STATE Selected agents $S^* \leftarrow \emptyset$ 
		  \STATE {Matching  $M \leftarrow \emptyset$}
	\FOR{agent $s\notin S^*$ down the list in $\succ_c$}
	\IF {$|S^*| < q_c$}
	\STATE Add $s$ to $S^*$ 
	\ENDIF
	\IF{ $|S^*| < q_c$ and if there exists an unmatched first rank seat $v_{t,i}^1$ for some type $t$ satisfied by $s$}
	\STATE Add  $(s,v_{t,i}^1)$ to $M$
	\ELSIF{ $|S^*| < q_c$ and if there exists a second rank seat $(s,v_{t,i}^2)$ for some type $t$ satisfied by $s$}
	\STATE Add  $(s,v_{t,i}^2)$ to $M$
	\ELSIF{$|S^*|<q_c$}
	 \STATE Add some new edge $(s,v_{t,i}^3)$ to $M$ where $t\in T(s)$.
	\ENDIF
	\ENDFOR
	       \RETURN $M$ and $S^*$.
\end{algorithmic}
\end{algorithm}

\begin{algorithm}[h!]
\begin{algorithmic}[scale=1]
	\REQUIRE $S'\subseteq S$, $q_c$, $\eta_c$, $\succ_c$.
    \ENSURE A matching $M\subseteq S'\times V$ and 
	a set of matched agents $S^*\subseteq S'$
	\end{algorithmic}
	\begin{algorithmic}[1]
	\caption{Priority Only Smart (POS)}
	\label{alg:pos}
\STATE Take top $q_c$ students $S^*$ with respect to $\succ_c$
\RETURN A-S applied to $(S^*$, $q_c$, $\eta_c$, $\succ_c$).
\end{algorithmic}
\end{algorithm}

%
\begin{example}[Priority Only Algorithms]
	\label{ex:PO}
	Consider the problem instance described in Example~\ref{ex:1}.
	Both priority based algorithms will select students $S^{*} = {s_1,s_2,s_3}$ filling no rank 1 seats and 2 rank 2 seats.
	The matching generated for both algorithms will be $\{(s_1,v^3_{t_0,1}),(s_2,v^{2}_{t_4,1}),(s_3,v^{2}_{t_3,1})\}$.
	Both algorithms simply select the highest ranked 3 students, but their main difference arises in the way in which they assign seats.
\end{example}

\section{Experimental Comparison}
We use two sets of synthetic data in order to compare our algorithms.
The first dataset is based on the SAT: the US university entrance examinations.
In this dataset, we generate data to match the relative diversity of test-takers in the US.
The goal of this dataset is to compare the performance of our selected matching algorithms in a real-world setting in which they can be utilised.

The first dataset is limited in testing scope by the total number of first and second rank reserves, which we will define as $\psi = $ ($\sum_{k=1}^{|types|} (|\eta_{t_{k}}^{1}|+|\eta_{t_{k}}^{2}|$), being less than $q_c$.
In order to overcome this limitation, for our second dataset, described in Section \ref{section:synthetic}, we generate data based on $\psi$ values exceeding $q_c$. 

\subsection{Comparison using synthetic SAT data}
\label{subsection:SAT}
In this section, we compare the performance of our selected algorithms when selecting a variable number of applicants from an input of 100 applicants with randomly generated types and priority ranking.
The diversity types we consider are "disadvantaged minority", "low parental education", and "low income household". 
These types are generated based on \citep{Coll20a}.
The "disadvantaged minority" type is an aggregation of the Black, Hispanic, and American Indian ethnicities, "low parental education" consists of applicants whose highest level of parental education is less than a bachelor's degree, and "low income household" applicants are those who used an SAT fee waiver.

For each selection capacity level, we simulate 100 datasets for consistency.
\subsubsection{Dataset generation}
In this dataset, we have a consistent number of applicants $|S| = 100$, and examine the performance of the algorithms across a range of different selection capacities $q_c$.
For each $q_c$, we generate 100 different applicant pools and aggregate the performance of the selected algorithms across all applicant pools.

For each applicant pool, we generate the diversity types and SAT score for every applicant individually.
We assign students' types with probabilities such that the total number of students with a given type matches the type frequencies outlined in \citep{Coll20a}.
When assigning overlapping types, we consider the conditional probability that students have a type given that they have already been assigned another type.
For example, disadvantaged minorities have a 1.7 times higher chance of coming from a low education household \citep{NiSc18a}.
We assign each applicant the ``disadvantaged minority'' type with a 39\% probability.
If the applicant is a disadvantaged minority, we assign him to be from a low education household with a 64\% probability, otherwise, we assign him to be from a low education household with a 30\% probability.
For applicants who have accrued both of the previous two types, we assign them to be from a low income household with a 30\% probability, otherwise, if they have only one type so far---this probability is 26\%, finally, if an applicant has no types so far, the probability of them being low income is 10\%.

We generate a student's SAT score using a truncated normal distribution. 
In this truncated normal distribution, the domain is 0 to 1600, the variance we use is 211 (as per \citep{Coll20a}), and the mean of the distribution is based on the types the applicant satisfies.
Students without any types have a mean score of 1135, while disadvantaged minorities score 172 points lower, applicants with low parental education score 171 points lower, and low income household applicants score 86 points lower on average.

We then reduce the expected score of each applicant in the truncated normal distribution based on the types they satisfy.
For students with multiple types, we reduce the impact of each type harmonically such that the impact of overlapping types is reduced.
For example, a student with all 3 types will have the expected value of their score reduced by 172 due to the disadvantaged minority type, $\lceil 171/2 \rceil = 86$ due to the low parental education type, and $\lceil 86/3 \rceil = 29$ due to the low income household type.
Having generated types and SAT scores for every student, we create a priority list $\succ_c$ by descending order of SAT scores.

The reserves $\eta_c$ were generated in a consistent manner proportional to the selection capacity $q_c$ for all datasets.
Writing the disadvantaged minority type as $t_1$, low education household type as $t_2$, and the low income household type as $t_3$, we have:
$\eta^{1}_{t_1} = 0.15 \times q_c$ , $\eta^{2}_{t_1} = 0.2 \times q_c$, $\eta^{1}_{t_2} = 0.1 \times q_c$,$\eta^{2}_{t_2} = 0.1 \times q_c$,$\eta^{1}_{t_3} = 0.05 \times q_c$, $\eta^{1}_{t_2} = 0.05 \times q_c$.
Recall that $\eta^{j}_{t}$ denotes the quota of type $t$ and rank $j$.
Hence, the total number of reserves is $\psi = 0.65q_c$ throughout this section.
\subsubsection{Performance of algorithms}
For each of the synthetic applicant pools generated above, we applied all of the algorithms outlined in Section~\ref{sec:toolbox}, then calculated performance metrics to compare their performance with respect to maximising diversity and selection of top performers.

For an instance $I$ and an algorithm $f$, let the outcome (selected students) of applying $f$ be $f(I)$. 
For an outcome $f(I)$, let $P(f(I))$ denote the performance of $f(I)$ with respect to a performance parameter $P$, where $P$ may be the number of rank 1 reserves filled, total reserves filled, or the average percentile rank of students.

For an algorithm $f$, performance measure $P$ and a set of instances $\mathcal{I}$, we define the average performance of an algorithm $f$ as $\text{avg}_{I\in \mathcal{I}} \{\frac{P(f(I))}{\text{opt}(P(I))}\}$ where $\text{opt}({P}({I}))$ denotes the maximum value of $P(I)$ reached by all algorithms for instance $I$. 
Informally, this calculates the ratio of the performance achieved by an algorithm $f$ for a metric $P$ relative to the best performance achieved by all algorithms for the same metric for each instance, and averages these ratios across all instances.

For an algorithm $f$, performance measure $P$ and a set of instances $\mathcal{I}$, we define the worst case perfomance of an algorithm $f$ as $\text{min}_{I\in \mathcal{I}} \{\frac{P(f(I))}{\text{opt}(P(I))}\}$. 
Informally, we calculate the same ratios as we do in the average case, but we take the lowest value of the calculated ratio across all instances rather than averaging them.

We define three main performance metrics by which we evaluate our matching algorithms. 
For a given algorithm $f$ and an outcome $f(I)$: 
\begin{enumerate}
	\item $P_{1}(f(I))$ denotes the number of first rank reserves satisfied by $f$,
	\item $P_{2}(f(I))$ denotes the total number of first and second rank reserves satisfied by $f$, and
	\item  $P_{3}(f(I))$ denotes the average percentile rank of students in $f(I)$.
\end{enumerate}

We present below the average and worst case performance of each algorithm relative to our performance metrics outlined above.

\begin{figure}[h!] 
    \centering
    \includegraphics[scale=0.4]{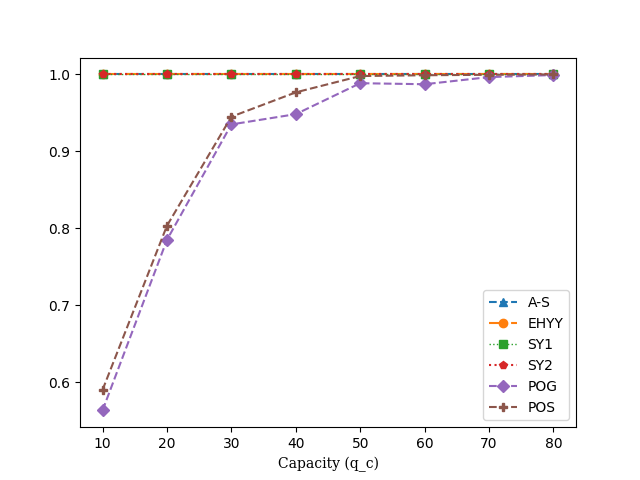}
    \caption{Average performance with respect to $P_1$}
    \label{fig:p1}
\end{figure}

In Figure~\ref{fig:p1} we see that the four diversity based algorithms (A-S, EHYY, SY1, SY2) are equivalent on average when selecting rank 1 seats, while the priority algorithms (POG, POS) trail behind.

\begin{figure}[h!] 
    \centering
    \includegraphics[scale=0.4]{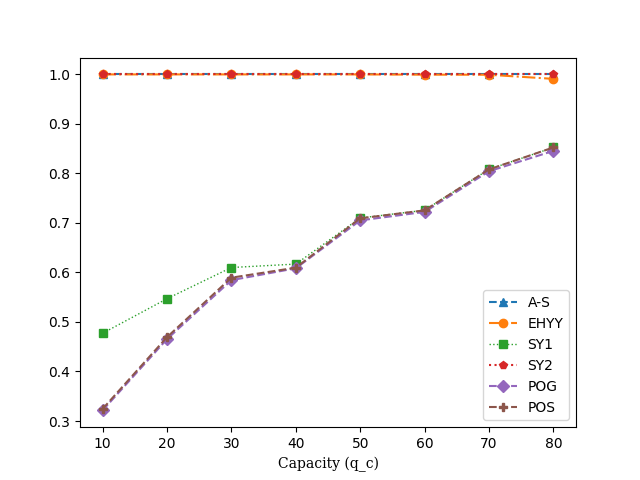}
    \caption{Average performance with respect to $P_2$}
    \label{fig:p2}
\end{figure}

In Figure~\ref{fig:p2}, we see that with respect to $P_2$, A-S and SY2 perform optimally, while EHYY is optimal up to higher values of $q_c$.
SY1 beats POG and POS (which overlap here) for lower values of $q_c$, yet converges with the priority algorithms at higher $q_c$ levels.

\begin{figure}[h!] 
    \centering
    \includegraphics[scale=0.4]{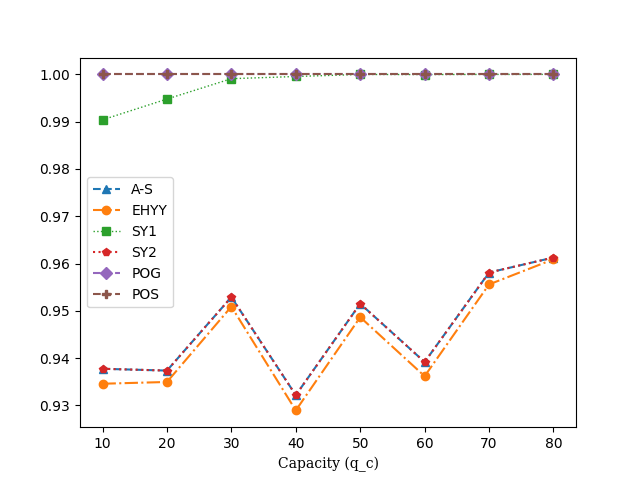}
    \caption{Average performance with respect to $P_3$}
    \label{fig:p3}
\end{figure}

In Figure~\ref{fig:p3}, POG and POS are optimal for all $q_c$ with respect to $P_3$, while SY1 trails closely behind.
A-S and SY2 overlap in performance, while EHYY exhibits the worst performance across the tested algorithms.

\begin{figure}[h!] 
    \centering
    \includegraphics[scale=0.4]{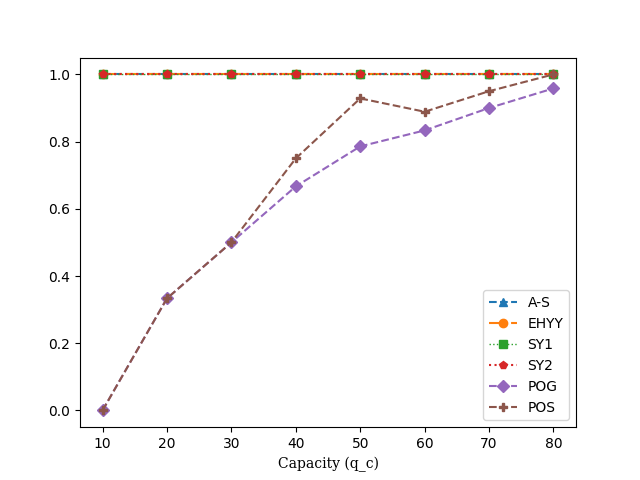}
    \caption{Worst case performance with respect to $P_1$}
    \label{fig:p4}
\end{figure}

In Figure~\ref{fig:p4}, the four diversity algorithms all overlap with equivalent performances.
POG and POS are equivalent for lower values of $q_c$ ($\leq 30$), while POS outperforms POG on higher values of $q_c$.

\begin{figure}[h!] 
    \centering
    \includegraphics[scale=0.4]{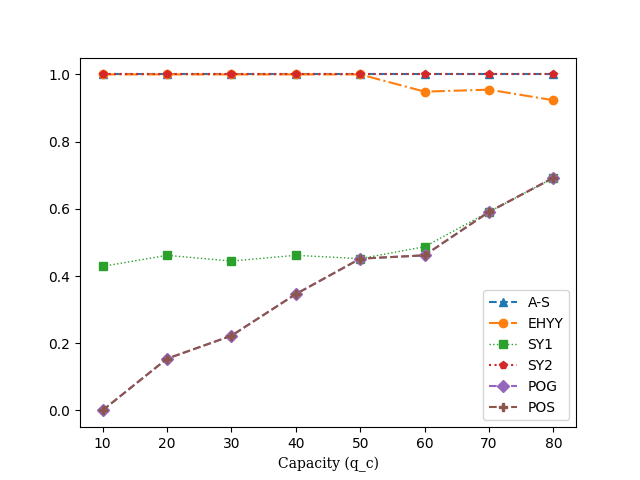}
    \caption{Worst case performance with respect to $P_2$}
    \label{fig:p5}
\end{figure}

In Figure~\ref{fig:p5}, A-S and SY2 perform equivalently, achieving an optimal result for all values of $q_c$.
EHYY is optimal for smaller $q_c$ values ($\leq 50$), however, its performance worsens for larger values.
SY1 outperforms the two priority algorithms at lower values, but converges to POG and POS at higher $q_c$.

\begin{figure}[h!] 
    \centering
    \includegraphics[scale=0.4]{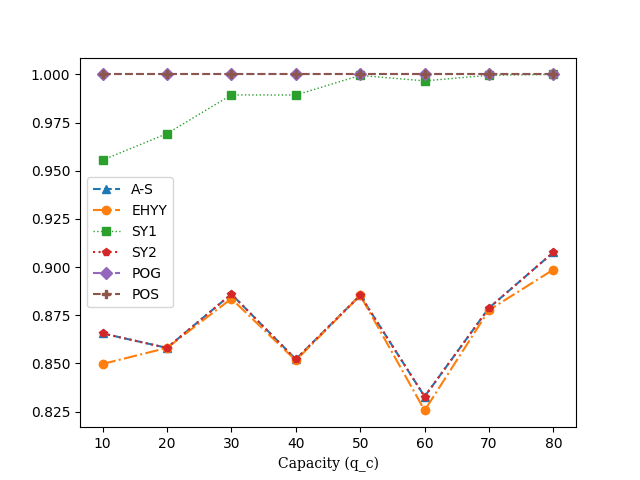}
    \caption{Worst case performance with respect to $P_3$}
    \label{fig:p6}
\end{figure}

In Figure~\ref{fig:p6}, POG and POS are optimal, SY1 clearly outperforms other 
diversity algorithms, and A-S, EHYY, and SY2 are largely equivalent, with EHYY marginally underperforming.

\subsubsection{Analysis}

By observing all of the figures above, we make the general observation that A-S and SY2 have an identical performance for the above dataset.
This is as $\psi$ is less than $q_c$ by a relatively large margin ($\psi = 0.65q_c$),
meaning that both SY2 and A-S are able to fill every reserve (regardless of rank) with the highest ranked students possible.

We also notice that POG and POS converge with SY1 for larger $q_c$ values. 
To explain this, we first note that since SY1 is unaware of rank 2 reserves, any difference between the two algorithms is purely based on rank 1 selection.
As we have distributed students with types lower than students without types on average, for a greater $q_c$, where priority algorithms will select lower ranked students, there will be a greater abundance of typed students, allowing POG and POS to achieve greater diversity results.
Since POG and POS both fill rank 1 seats before rank 2 seats (after selection), they improve on their first rank diversity before improving on second rank diversity as $q_c$ increases.
Hence, POG and POS approach SY1 quickly, before filling rank 2 seats to approach the other diversity algorithms.

Figures \ref{fig:p1} and \ref{fig:p4} show that all of the diversity based algorithms are equivalent with respect to one rank ($P_1$),
while the performance by POG and POS improves drastically as $q_c$ increases, approaching optimality.

Figures \ref{fig:p2} and \ref{fig:p5} show that A-S and SY2 are optimal with respect to the first two ranks, while EHYY is optimal for lower values of $q_c$.
We note that, since SY1 is unaware of rank 2 seats, the performance gap between SY1 and the other diversity algorithms is entirely made up of the number of rank 2 reserves filled.

From figures \ref{fig:p3} and \ref{fig:p6} we see that SY1 outperforms other diversity based algorithms as a result of satisfying fewer reserves (and hence satisfying a larger proportion of $q_c$ based only on priority).
We also notice that SY2 and A-S marginally outperform EHYY for all values of $q_c$, as the greedy selection approach by EHYY does not ensure optimality with respect to priority.

From this dataset (where the number of rank 1 and 2 reserves is less than $q_c$) we find that:
\begin{enumerate}
	\item{When maximising diversity with respect to one rank, all diversity algorithms are optimal.}
	\item{When maximising diversity with respect to the first two ranks, A-S, SY2, and EHYY produce near-identical results.}
	\item{For $q_c \geq 30$, priority based algorithms fill more than 90\% of the optimal number of rank 1 seats.}
	\item{For $q_c \geq 70$, priority based algorithms fill more than 80\% of the optimal number of rank 1 and 2 seats.}
\end{enumerate}

\subsection{Comparison using random synthetic data}
\label{section:synthetic}
In our comparison of the algorithms using the synthetic SAT data above, we have been limited in scope by keeping a mostly consistent set of paramaters in order to simulate a student admission problem.
In this section, we explore scenarios in which $\psi > q_c$.
We thus vary both the number of total (rank 1 + rank 2) reserves available for agents, as well as our acceptance capacity ($q_c$) to further compare the performance of our matching algorithms. 

\subsubsection{Data procurement}
We maintain the same testing conditions as used in the SAT data above, but with the key difference of varying $\psi$ across different snapshots, while varying $q_c$ within each snapshot. 
We choose our three main values for $\psi$ as 1.3, 1.5 and 1.7 times of $q_c$, which we will achieve by multiplying the quotas found in Section~\ref{subsection:SAT}.
For example, instances with $\psi = 1.3q_c$ will have double the number of reserves of each type and rank than the corresponding instances in Section~\ref{subsection:SAT}, as the instances in Section~\ref{subsection:SAT} have $\psi = 0.65$ $q_c$. 
For each of these three values, we compare our selection algorithms across four values of $q_c$, namely 20, 40, 60, and 80 with a consistent $|S| = 100$.
When evaluating the performance of our algorithms, we will use the same metrics for comparison of the algorithms as Section \ref{subsection:SAT}.
While we have done average and worst case testing for $\psi = \{1.3q_c,1.5q_c,1.7q_c\}$, we only include the worst case graphs for $\psi = 1.7q_c$ for the sake of space.
\subsubsection{Performance of algorithms for $\psi = 1.7$}

\begin{figure}[h!] 
    \centering
    \includegraphics[scale=0.4]{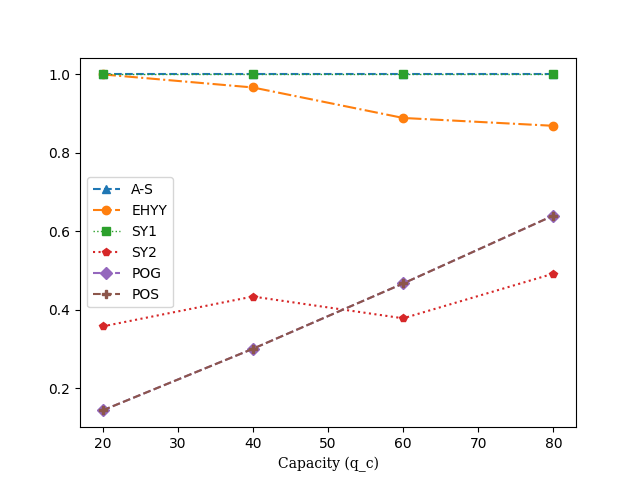}
    \caption{Worst case performance for $P_1$, $\psi = 1.7q_c$}
    \label{fig:p22}
\end{figure}

In Figure~\ref{fig:p22}, A-S and SY1 overlap at $P_1 = 1$. The next best algorithm is EHYY which trends downwards as $q_c$ increases.
SY2 outperforms POG and POS (which overlap) at lower $q_c$, but is overtaken at higher values of $q_c$.

\begin{figure}[h!] 
    \centering
    \includegraphics[scale=0.4]{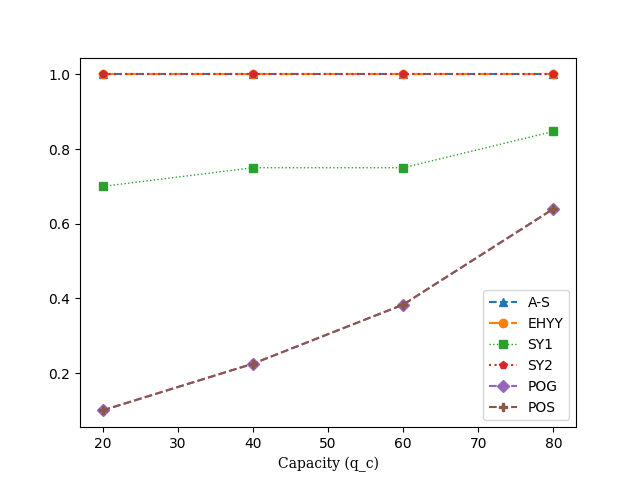}
    \caption{Worst case performance for $P_2$, $\psi = 1.7q_c$}
    \label{fig:p23}
\end{figure}

In Figure~\ref{fig:p23}, A-S, SY2, and EHYY overlap at $P_2 = 1$. SY1 then heavily outperforms POG and POS, which are overlapping.

\begin{figure}[h!] 
    \centering
    \includegraphics[scale=0.4]{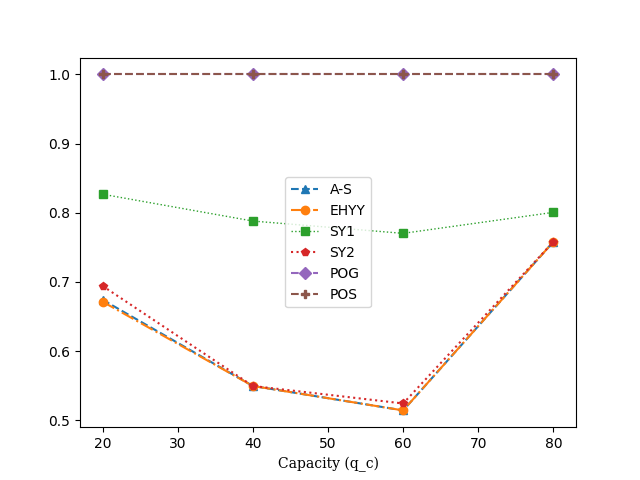}
    \caption{Worst case performance for $P_3$, $\psi = 1.7q_c$}
    \label{fig:p24}
\end{figure}

In Figure~\ref{fig:p24}, POG and POS overlap at $P_3 = 1$. SY1 is the next best performing, then SY2 outperforms EHYY and A-S, which overlap.

\subsubsection{Analysis}
We see significantly different performance from each algorithm compared to what has been demonstrated in Section \ref{subsection:SAT}, as well as between different $\psi$ levels.
The most notable difference from Section \ref{subsection:SAT} is that, for instances where $\psi > 1$, A-S and SY2 are no longer equivalent, but POG and POS are equivalent.

From observing performance relative to $P_1$, we have gained the following results.
\begin{enumerate}
	\item A-S and SY1 remain optimal for $P_1$.
	\item EHYY's $P_1$ performance is close to 1 for all $q_c$ in the average case, but drops drastically in the worst case, especially at higher values of $\psi$ and $q_c$.
	\item SY2 outperforms POG and POS in satisfying $P_1$ for lower $q_c$ values, but is overtaken for higher $q_c$.
\end{enumerate}

Relative to $P_2$, we obtain the following results.
\begin{enumerate}
	\item A-S, SY2, and EHYY are optimal for $P_2$.
	\item SY1 is strictly better than POG and POS when satisfying $P_2$.
	\item The performance of SY1 increases as $\psi$ increases, both in terms of average and worst case.
\end{enumerate}

By comparing $P_3$ performances, we obtain the following results.
\begin{enumerate}
	\item POG and POS remain optimal for $P_3$.
	\item SY1 is significantly better at satisfying $P_3$ than other diversity based algorithms.
	\item The performance of all diversity based algorithms decreases relative to $P_3$ as $\psi$ increases.
	\item Diversity based algorithms perform worst at $P_3$ for intermediate levels of $q_c$ that is, $q_c = {40,60}$.
\end{enumerate}

\section{Conclusions}
We have examined the effectiveness of prominent matching algorithms in satisfying a range of performance metrics across a variety of different instances.
We find that there is a necessary tradeoff when balancing performance between priority and reserves, and this tradeoff can be negotiated through our choice of selection algorithm.

When we wish to optimise our matching toward fulfilling reserves across multiple ranks, the A-S algorithm will always provide the best solution while maintaining the highest possible priority of selected agents.
However, if we wish to optimise across only one rank, SY1 and SY2 can provide a solution that can achieve this while outperforming A-S in terms of priority ranking.
It also becomes clear that, for most instances where $q_c$ is not high, reserve based matching algorithms provide highly different outcomes from priority-only algorithms such as POG and POS, creating further emphasis on the tradeoff between priority and reserve satisfaction.

Therefore, when selecting an algorithm to solve a problem, we must carefully consider the following points:
\begin{enumerate}
	\item Whether or not the problem requires optimisation for priority or reserves.
	\item The relative importance of filling reserves according to rank against the importance of maximising priority.
	\item The value of {capacity} $q_c$ relative to {the number of students} $|S|$.
	\item The number of reserves available relative to  {capacity} $q_c$.
\end{enumerate}

%
%
		



\bibliographystyle{ACM-Reference-Format}  

\bibliography{abb,adt}

\end{document}